\def\thin{{\thinspace}}

\def\ref{\par \noindent \hangindent=3pc \hangafter=1}

\def\sep{{\par \noindent \hangindent=15pt \hangafter=1}}

\def\scr{\scriptstyle}
\def\etal{{{\it et al}.\ }}

\def\pn{{\par\noindent}}
\def\ie{{{\it i.e}.}}

\def\rbar{{\overline{r}}}
\def\rhobar{\lower 0.2ex\hbox{${\overline{\rho}}$}}

\def\;{;\thin}

\def\vecOmega{{\rlap{$\Omega$}{\hskip 0.1ex\hbox{$\Omega$}}}}

\def\vectheta{{\rlap{$\theta$}{\hskip 0.1ex\hbox{$\theta$}}}}
\def\thetahat{{\rlap{\vectheta}{\hskip 0.2ex{\raise 0.5ex\hbox{\overline}}}}     }

\def\r{{{\bf r}}}

\def\w{{\ \ \ \ \ }}

\def\newpage{{\vfill\eject}}

\def\tgs{{\thin \rlap{\raise 0.5ex\hbox{$\scr  {>}$}}{\lower 0.3ex\hbox{$\scr  {\sim}$}} \thin }}
\def\tls{{\thin \rlap{\raise 0.5ex\hbox{$\scr  {<}$}}{\lower 0.3ex\hbox{$\scr  {\sim}$}} \thin }}
\def\tll{{\raise 0.3ex\hbox{$\scr  {\thin \ll \thin }$}}}
\def\tgg{{\raise 0.3ex\hbox{$\scr  {\thin \gg \thin }$}}}
\def\tle{{\raise 0.3ex\hbox{$\scr  {\thin \le \thin }$}}}
\def\tge{{\raise 0.3ex\hbox{$\scr  {\thin \ge \thin }$}}}
\def\tl{{\raise 0.3ex\hbox{$\scr  {\thin < \thin }$}}}
\def\tg{{\raise 0.3ex\hbox{$\scr  {\thin > \thin }$}}}
\def\ts{{\raise 0.3ex\hbox{$\scr  {\thin \sim \thin }$}}}

\def\do{{\tt "}}
\def\tp{{\raise 0.3ex\hbox{\fiverm +}}}
\def\Chi{{\raise 0.4ex\hbox{$\chi$}}}

\def\deg{{^\circ}}

\def\K{{\rm\thin K}}

\def\Msun{\hbox{$\thin M_{\odot}$}}

\def\Lsun{\hbox{$\thin L_{\odot}$}}

\def\Rsun{\hbox{$\thin R_{\odot}$}}

\def\cc{$\scr ^c$}
\def\dd{$\scr^d$}
\def\p{\thin\tp\thin}

\input psfig.sty
\documentclass[12pt,preprint]{aastex}
\newcommand{\be}{\begin{equation}}
\newcommand{\ee}{\end{equation}}
\begin{document}
\shorttitle{Envelope Ejection}
\title{Envelope Ejection: an Alternative Evolutionary Process for some Early Case B Binaries }
\author{Peter P. Eggleton\altaffilmark{1,2}}
\altaffiltext{1}{Lawrence Livermore National Laboratory, Livermore, CA 94550\\
    Email: {\tt ppe@igpp.ucllnl.org}}
\altaffiltext{2}{On leave from the Institute of Astronomy, Madingley Rd,
         Cambridge CB3 0HA, UK}
\def\rbar{{{r_*}}}
\def\k{{\bf k}}
\def\kdot{{\hbox{$\dot k$}}}
\def\kbfdot{{\hbox{$\dot {\bf k}$}}}
\def\r{{{\bf r}}}
\def\EEdot{\hbox{$\dot E$}}
\def\vbf{{{\bf v}}}
\def\vO{\vecOmega}
\def\et{equilibrium-tide\ }
\def\tf{tidal friction\ }
\def\smc{0045-7319}
\def\Te{T_{\rm E}}
\def\spc{\hskip 0.3truein}
\def\p{^{\prime}}
\def\olp{\omega_{\rm lp}}
\def\olpA{\omega_{\rm lp,A}}
\def\K{{\bf K}}
\def\H{{\bf H}}
\def\E{{\bf E}}
\def\Q{{\bf Q}}
\def\Jhat{\overline{\bf J}} 
\def\uhat{{\overline{\bf u}}}
\def\Hbar{{\overline{H}}}
\def\alpdot{{d\aJ\over dt}}
\def\olpdot{{\hbox{$\dot \omega_{\rm lp}$}}}
\def\tTF{t_{\rm F}}
\def\tTFa{t_{\rm F1}}
\def\Oe{\Omega_{\rm e}}
\def\Oq{\Omega_{\rm q}}
\def\Oh{\Omega_{\rm h}}
\def\Oae{\Omega_{\rm 1e}}
\def\Oaq{\Omega_{\rm 1q}}
\def\Oah{\Omega_{\rm 1h}}
\def\Obe{\Omega_{\rm 2e}}
\def\Obq{\Omega_{\rm 2q}}
\def\Obh{\Omega_{\rm 2h}}
\def\He{\overline{H}_{\rm e}}
\def\Hq{\overline{H}_{\rm q}}
\def\Hh{\overline{H}_{\rm h}}
\def\vr{{V_{\rm rot}}}
\def\aa{$^a$}
\def\bb{$^b$}
\def\cc{$^c$}
\def\dd{$^d$}
\def\PZ{P_{\rm ZAMS}}
\def\rz{R_{\rm ZAMS}}
\begin{abstract} 
Abstract:  We discuss the evolution of binaries with moderately high masses ($\ts 10 - 30\Msun$),
and with periods of $\ts 3 - 300\thin$d, corresponding mostly to early Case B. These are usually
thought to evolve either by reasonably conservative Roche-lobe overflow, if the initial mass 
ratio is fairly mild, or else by highly non-conservative common-envelope evolution, with spiral-in 
to short periods ($\ts$ hours, typically), if the initial mass ratio is rather extreme. We discuss 
here a handful of binaries from part of this period range $(\ts 50 - 250\thin$d), which appear 
to have followed a different path: we argue that
they must have lost a large proportion of initial mass ($\ts 70 - 80\%$), but without shortening
their periods at all. We suggest that their behaviour may be due to the fact that stars 
of such masses, when evolved also to rather large radii, are not far from the Humphreys-Davidson 
limit where single stars lose their envelopes spontaneously in P Cygni winds, and so have envelopes 
which are only lightly bound to the core. These envelopes therefore
may be relatively easily dissipated by the perturbing effect of a companion. In addition, some or
all of the stars considered here may have been close to the Cepheid instability strip when they
filled their Roche lobes. One or other, or both, of high luminosity and Cepheid instability, 
in combination with an appropriately close binary companion, may be implicated.
\end{abstract}
\section{Introduction}
\par It is well known that stars of high mass, $\tgs 30\Msun$, appear to be strongly affected by
mass loss at some stage in their evolution across the Hertzsprung-Russell diagram (HRD). Humphreys
\& Davidson (1979) found that stars are absent above a line (the Humphreys-Davidson limit)
in the HRD that slopes down gently from left to right. Theoretical evolutionary tracks for stars
$\tgs 30\Msun$ should cross this line, but apparently real stars do not. Furthermore stars near
the Humphreys-Davidson limit are often highly variable, and have indications (P Cygni line profiles,
variable light-curves) of fast, copious and erratic stellar winds. Thus it is likely that 
as a massive star approaches the Humphreys-Davidson limit it loses considerable
mass and instead of evolving further to the red is stripped almost to its helium-burning core.
It then evolves to a small hot remnant, a Wolf-Rayet (WR) star.
\par Evolutionary tracks of stars of lower mass do not intersect the Humphreys-Davidson limit. 
This does not mean that they suffer no mass loss at all, but it may mean that they do not suffer 
much mass loss until after they have crossed the HRD and become red supergiants. One can hope for 
guidance here from observed masses of late supergiants in binaries; but there are not many of 
these, and the difficulty of measuring the small radial-velocity amplitudes of the hot, and 
therefore typically broad-lined, component makes some determinations quite uncertain. We discuss 
some possibly relevant systems below.
\par Conservative early Case B evolution, starting from a mass ratio which is not 
strongly different from unity, is expected to pass through (i) Roche-lobe overflow 
(RLOF), then (ii) a detached phase, with the loser having become small, and (iii) 
a late stage of {\it reversed} Roche-lobe overflow (RLOF), as the gainer's evolution 
proceeds. We distinguish two main subtypes of Case B, depending on whether the loser 
is able to complete its evolution (presumably to a supernova explosion, followed by 
a neutron star) before the gainer evolves to RLOF, or on the other hand is still in 
a helium-burning stage when this happens. Following the discussion of Nelson \& 
Eggleton (2001), hereinafter Paper I, on Case A evolution, we call these subtypes 
Case BN (`no overtaking') and Case BL (`late overtaking'). In Section 2 we discuss 
three observed systems near the Case B/A borderline, that seem to be reasonable 
examples of the conservative model during the first RLOF phase, and three more systems 
that appear to be reasonable examples of the detached phase that follows. In Section
2 we also emphasise that there at least three other variants of Case B, which we call
Cases BB, BR and BD, by analogy with three subtypes of Case A in Paper I. 
In Section 3 we discuss four observed binary systems that we 
believe are highly evolved and which do not fit well the usual assumptions of Case B. 
Four is not a large number, but all four appear to us to show strong evidence that a 
considerable amount of mass has been lost from each system, and yet the orbital 
periods ($\ts 50 - 225\thin$d) are by no means as small as one would expect if the 
mass-loss process was driven primarily by the release of orbital energy during a 
common-envelope phase of evolution. We call this unexpected variant Case BU. In 
Section 4 we discuss a possible physical mechanism, and in Section 5 we discuss some 
implications for the evolution of binary stars.

\section{Some Examples of Case A and early Case B Evolution}

\par Paper I discussed the extent of agreement between a sample of relatively
well-determined Algol binaries and a grid of 5550 computed binaries undergoing 
conservative RLOF in Case A. A major result was that `hot Algols', those with 
{\it both} components earlier than $\ts $G0, appeared to agree reasonably well 
with the computed models, with the exception of two ($\lambda$ Tau, DM Per) that 
are known to have quite unusually close third bodies; these third bodies might 
influence the orbit of the close pair, by extracting angular momentum from it 
on a tidal-friction timescale (Kiseleva \etal 1998). On the other hand `cool 
Algols', in which one component is of type G or later, did not fit at all well.
This was attributed to the possibility of dynamo activity in cool convective
envelopes, which can lead to angular momentum loss by way of stellar winds,
magnetic braking and tidal friction. The discrepancies in the cool Algols all
appeared to be in the sense that the system had less angular momentum, and in 
some cases less mass, than the otherwise best-fitting theoretical models. Some
of these discrepancies could be accounted for quite well with a simplistic model
of dynamo activity, magnetically-driven mass loss, magnetic braking and tidal
friction described by Eggleton \& Kiseleva-Eggleton (2002; Paper II)
\par The present paper is preliminary to a similar study that we hope to make
regarding Case B. But the point we try to make here is that even without such
a substantial grid (and it would have to be considerably more substantial since
the range of periods is some 10 -- 100 times greater), it is reasonably clear 
both that some systems fit the theoretical paradigm reasonably well and that 
others cannot possibly do so. We hope to make a provisional judgment of what 
kind of evolutionary processes are required to produce overall agreement.
\par In Table 1 we list observed parameters for a number of systems which are 
relevant to our discussion, and the references from which the data were taken. 
The first three are systems which we believe fit 
fairly satisfactorily into the standard picture of current conservative RLOF. 
They were chosen to have a total mass in roughly the range we would like, 
and angular momentum corresponding to very early Case B, and in some cases
marginally Case A. The next three systems have substantially longer periods and 
arguably represent reasonably well the later stage of detached evolution beyond 
the point where the loser
has retreated inside its Roche lobe after extensive RLOF. The next four are, we 
believe, difficult systems which we leave to Section 3. The last six are 
some further systems, mostly wider than the first ten, which are relevant to 
the discussion in Section~4. 
\par The errors in the values quoted are discussed in the 
references. The better values -- normally from double-lined eclipsing systems -- 
have uncertainties that are fairly typically of order $\pm 10\%$, but in some 
other cases there is an element of assumption or supposition whose magnitude is 
difficult to assess.
\par Note that we use a convention that suffices 1 and 2 refer to the components that
were {\it initially} more and less massive, respectively. This is not a standard
convention, but since we are discussing systems whose component masses, temperatures and
luminosities may all change considerably during evolution we feel that it is more
logical to use suffices that do not change in the course of evolution. There is of
course an element of supposition, but we believe that Table 1 makes clear what
supposition we are using, and we attempt to justify these suppositions in the text.
Under `spectra' in Table 1, the two spectra are listed in this suppositional order,
with a question mark for one or other component if it is not seen, or is seen but
not classified.
\par We furthermore use a convention that the mass ratio $q$ is {\it always}
$M_1/M_2$, and not its reciprocal. Thus $q\tg 1$ at age zero; it may drop below
unity during RLOF, and might in principal rise above it again in the later
reversed-RLOF phase (although we shall argue that in practice this is unlikely).
\begin{table}
\caption{System parameters for some binaries}
\begin{center}
\begin{tabular}{clllllllllll}
\tableline\tableline
name &alias&spectra &$P$&$e$&$M_1$ &$M_2$&$R_1$ &$R_2$ &$X$\aa&$Y$\aa&reference\\
\tableline 
V356 Sgr\dd&HD173787        &A2II + B3         &8.90  &    &4.7    &12.1     &14     &6          &6.3    &1.36    & 1\\
AQ Cas\dd  &BD+61$\deg$0242 &B9 + B3           &11.7  &    &4.7    &16.6     &16.6   &7.6        &4.7    &1.42    & 2\\
RZ Sct\dd  &HD169753        &F5 + B3II         &15.2  &    &2.5    &11.7     &15.9   &15.8       &4.3    &3.7     & 3\\
&&&&&&&&&&&\\
$\phi$ Per     &HR496       &HeIem +B1IIIpe    &127   &    &1.15   &9.3      &1.3    &$5.5-8$\cc &12     &1.8:    & 4\\
3 Pup          &HR2996      &? + A2Iabe        &161   &    &       &.006\bb  &       &           &       &        & 5\\
               &HD51956     &B2-3e + F8Ib      &107   &    &       &.0016\bb &       &           &       &        & 6\\
&&&&&&&&&&&\\
V379 Cep\dd    &HR7940      &B2III + B2III     &99.7  &.15 &1.9    &2.9      &5.2    &7.4        &       &        & 7\\
$\upsilon$ Sgr &HR7342      &AIp + ?           &138   &    &2.5    &4.0      &       &           &       &        & 8\\
               &0045-7319   &NS(.926s) + B1V   &51.2  &.81 &1.4    &8.8      &       &6.4        &       &1.75    & 9\\
V2174 Cyg      &HD235679    &BN2.5Ibe + ?      &225   &.1  &5.9\bb &         &       &           &       &        & 10\\
&&&&&&&&&&&\\
S Mus          &HR4645      &F6Ib + B3.5V      &505   &.08 &.165\bb&         &       &           &       &        & 11 \\
AX Mon         &HD45910     &K0-4III + B2pe    &233   &    &4.5    &13.6     &120    &14         &127    &3.0     & 12 \\ 
V695 Cyg\dd    &HR7735      &K4Ib + B4V        &3784  &.22 &7.2    &5.5      &170    &4.0        &5300   &1.44    & 13 \\
V1488 Cyg\dd   &HR7751      &K5Iab + B7V       &1145  &.30 &7.2    &4.1      &170    &3.1        &1800   &1.32    & 13 \\
VV Cep         &HR8383      &M2epIab + B0      &7430  &.35 &20     &20       &1600   &13         &6300   &2.2     & 14 \\
AZ Cas      &BD+60$\deg$310 &M0epIab + B0V     &3404  &.55 &18     &13       &450    &30         &3400   &6.5     & 15 \\ 
\tablecomments{ Periods in days; eccentricities are zero unless otherwise stated; masses and radii
are in Solar units.
\pn \aa See text for definition
\pn \bb Mass function
\pn \cc Polar -- equatorial radii
\pn \dd Double-lined eclipsing system.
\pn References: 1 -- Popper 1980; 2 -- Olson 1994; 3 -- Olson \& Etzel 1994;  4 -- Gies \etal 1998; 
\pn 5 -- Plets \etal 1995; 6 -- Burki \& Mayor 1983, Ake \& Parsons 1990; 7 -- Gordon \etal 1998; 
\pn 8 -- Dudley \& Jeffery 1990; 6 -- Kaspi \etal 1994, 1996, Bell \etal 1995; 10 -- Bolton \& Rogers 1978;
\pn 11 -- Evans 1990; 12 -- Elias \etal 1997; 13 -- Schr{\"o}der \etal 1997; 14 -- Wright 1977; 15 --  Cowley \etal 1977}
\end{tabular}
\end{center}
\end{table}
\par The evolutionary code used here differs slightly from that used in Paper I; it has
been described in Paper II. This code includes tidal
friction, and the interaction of the intrinsic spin of star 1 with orbital eccentricity and 
orbital angular momentum. The effect of the spin of star 1 on its internal structure is
included, though only as {\it uniform} rotation. For present purposes this makes little 
difference. The code does {\it not} include the intrinsic spin of star 2, for reasons stated 
in Paper II. This may be more important, as we see in some of the systems discussed below, 
and we would hope to rectify this in the future.
\par If an Algol system has evolved conservatively from zero age, then the quantity 
$PM_1^3M_2^3$ should have been constant, as well as $M\equiv M_1+M_2$. $P$ 
therefore has a minimum at $M_1=M_2$,
and in practice changes rather little ($\tls 12\%$) over the mass-ratio range 
$1.5\tge q\tge 0.67$. We find it helpful to think in terms of a canonical zero-age
mass ratio $q_c=4/3$, which in turn defines a canonical zero-age mass $M_{1c}$,
supposing that we know both masses currently. From this we can define a critical
period  $\PZ$, which is the orbital period of a binary containing a zero-age main 
sequence (ZAMS) star of mass $M_{1c}$ that just fills its Roche lobe. Case A 
evolution occurs if the initial period $P_0$ lies in roughly the range 
$1\tl X\equiv P_0/\PZ\tls 2.5-8$ (Paper I; the upper limit depends on the mass), 
and early Case B for larger values
up to a limit determined by either the onset of He burning or the development
of a deep convective envelope. The critical ratio is $\ts 5$ at the fairly high
masses considered here, but ranges from $\ts 2.5$ to $\ts 6$ as $M_{1c}$ ranges
from $\ts 1$ to $\ts 30\Msun$.  We anticipate that
most systems with $5\tls X\tls 100$ will be early Case B systems. The above
prescription defines the parameter $X$ which is tabulated in a few cases in
Table 1. To summarise,
$$M_{1c}={4\over7}(M_1+M_2)\ ,
\w M_{2c}={3\over 7}(M_1+M_2)\ ,\eqno(1)$$ 
$$X\equiv{PM_1^3M_2^3\over \PZ(M_{1c})M_{1c}^3M_{2c}^3}\ ,\eqno(2)$$
$$  \PZ(M) \  \approx\ 0.3616\thin\sqrt{\rz^3(M)\over M}\ ,\eqno(3)$$
$$ \rz(M) = {1.715 M^{2.5} + 6.60 M^{6.5} + 10.09M^{11}  + 1.0125 M^{19} + 0.0749 M^{19.5}\over
0.01077 + 3.082 M^2 + 17.85 M^{8.5} +  M^{18.5} + 2.26\times 10^{-4}M^{19.5}} \ \ . \eqno(4)$$
We use the approximation of Tout \etal (1996) for the ZAMS radius in the range 
$0.1\tls M\tls 250\Msun$, assuming solar metallicity.
\par The suffix $c$ stands for `canonical starting value': we do not suppose that
all systems start with exactly this mass ratio (4/3), but $X$ is a useful indicator
of Case A or Case B provided that the actual initial masses $M_{10}, M_{20}$ are
not very different from the canonical starting values. We note that for V356 Sgr 
$X=6.3$, suggesting very early Case B, and for AQ Cas and RZ Sct the slightly 
lower values are close to the Case A/B borderline.
\par For systems with known masses {\it and} radii, i.e. usually eclipsing
double-lined systems in favorable circumstances, a second useful parameter $Y$
is the ratio of the current radius of star 2 to the ZAMS radius that it would
have at its current mass:
$$Y\equiv {R_2\over \rz(M_2)}\ .\eqno(5)$$
$Y$ is indicative of the amount of
evolution that has gone on within the star, both before and after RLOF.
It seems unlikely that $Y$ could be less than unity, and in practice this
seems to be true of the great majority of systems that we have checked (but
do not list here). However $Y$ can be greater than unity on account of
both {\it nuclear} expansion before and after RLOF begins, and 
{\it thermal} expansion during the more rapid early phase of the RLOF. 
Possibly $Y$ can also be large on account of the spin angular momentum that
star 2 acquires during RLOF, which we discuss further below but do not model 
here. But $Y$ can also {\it decrease} during mass transfer, because as a main-
sequence star gains mass its convective core grows disproportionately faster,
and so dilutes the concentration of burnt fuel in the core.
\subsection{V356 Sgr}
\par For V356 Sgr the modest value $Y\ts 1.36$ is actually quite a challenge to 
theoretical models: (a) if the initial mass ratio was fairly close to unity star 
2 should have evolved quite significantly during the pre-RLOF phase, while 
(b) if it was fairly far from unity then the rather mild current 
mass ratio ($q\ts 0.4$) would mean that thermal-timescale mass transfer is still
continuing. Either of these could give a value of $Y$ substantially above
what is observed. In a tentative exploration of parameter space we find that
starting values ($9.4+7.4\Msun;5\thin$d) give something like the present parameters
at age $26\thin$Myr, but with $R_2\ts 7.9\Rsun$ rather than $6\Rsun$. Since
Popper (1980) lists the uncertainties of both the masses and the radii as
$\ts 10\%$, we feel that the discrepancy is modest and might be eliminated in 
a careful search of parameter space, with full allowance for observational
uncertainty.
\par The evolution of the two components is shown in Fig 1: (a) the theoretical 
HRD, (b) the stellar and Roche-lobe radii as functions of mass ratio, and (c) 
the stellar radii as functions of age. The evolution of star 1 was terminated 
automatically when either (i) it was in the process of carbon ignition, presumably 
followed by further (short-lasting) core carbon burning and a supernova explosion, 
(ii) when star 2 filled its own Roche lobe, or (iii) after 3000 timesteps. 
Carbon ignition applied in this case. Immediately before this star 1 was a shell 
helium-burning star (but with a thin hydrogen-rich envelope)
of mass $2.24\Msun$, expanding rapidly back towards the giant branch but
liable to explode well before then. The result would no doubt be a high-mass
X-ray binary (HMXB), unless there is a sufficient asymmetric kick to disrupt 
the system. The orbital period increased to $48.3\thin$d. 
\par We find it convenient to subdivide Case B into five subtypes, BN, BL,
BB, BR and BD, as follows:
\sep BN -- No overtaking -- star 1 goes through the whole of its evolution
to a compact remnant (WD, NS or BH) before star 2 in turn fills its Roche lobe
\sep BL -- Late overtaking -- star 2 reaches its Roche lobe after star 1 has
detached from its Roche lobe, but before star 1 becomes a compact remnant
\sep BB -- classic Case BB$^{\prime}$ -- star 1 has a second phase of RLOF, as 
a helium red giant, before contracting or exploding to a compact remnant
\sep BR -- Rapid evolution to contact -- star 2 reaches its Roche lobe during
its rapid (thermal) expansion shortly after star 1's RLOF begins
\sep BD -- Dynamic evolution, probably to a common-envelope phase -- at a
large initial mass ratio, star~1 may not be able to contract as fast as its Roche
lobe, and so the mass transfer climbs to a very high (hydrodynamical) rate.
Probably star~2 also expands very rapidly, and the system goes into deep
contact.
\pn Paper I, although mainly concerned with Case A, had examples of all of
these Case B subtypes at periods slightly above the Case A boundary. Our model of 
V356 Sgr is fairly typical of Case BN, but is rather close to Case BB.  
\par Cases BR and BD require rather large initial mass ratios $q_0$, with the
required minimum $q_0$ itself a fairly strong function of $X$. The wider the
initial binary, the better chance star 2 has of expanding substantially and yet
failing to fill its own Roche lobe even when its Roche-lobe radius is a
minimum at about equal masses -- see RZ Sct below. We estimate that Case BR
needs $q_0\tgs 1.6$ at about the A/B borderline. Case BD requires a larger
$q_0$ still, possibly $q_0\tgs 3$; but we have not reached a clear estimate
of this critical value since we find it rather hard to define clearly the
boundary between thermal and hydrodynamic rates of mass loss.
\par Note that although we expect, and find, the above 5 subcases of Case B
in theoretical models of conservative Roche-lobe overflow, our conclusion below 
(Section 3) is that at least in some cases, apparently rather wider systems 
than the first three in Table 1, the overflow is markedly non-conservative.
We may therefore have to add at least one more subcase to the above list.
\subsection{AQ Cas}
\par AQ Cas can be modeled slightly more accurately than V356 Sgr, partly because 
$Y$ is a little larger, partly because the current $q$ is a little smaller, and 
partly because the system has less angular momentum, corresponding marginally to Case A. 
Starting conditions ($12 + 9.3\Msun, 4\thin$d) gave a radius for star~2 of 
$6.5\Rsun$ when the present masses and period were reached; slightly {\it smaller} 
than required, but probably within experimental error. Its evolution is shown
in Fig 2, and is Case AN, as defined in Paper I (but analogous to Case BN above).
Because RLOF begins before the normal `hook' in the HRD is reached, a modified
hook appears just below the middle right: a brief detached phase interrupts the RLOF.
This occured at parameters $5.6 + 12.6\Msun; 8.0\thin$d. The evolution terminated
itself as (almost) non-degenerate carbon ignition got under way. Star 1 was heading
back towards its Roche lobe, but was not very close to it.
\subsection{RZ Sct}
\par RZ Sct poses a greater problem than the two previous systems: its 
{\it large} value of $Y$ is very challenging. It argues for either {\it very} 
closely equal initial masses, so that star 2 evolves almost as much as star 1 
before RLOF and then further once its mass starts to increase, or else for 
rather different masses (perhaps $q_0\tgs 2$) so that RLOF is still in the 
thermal phase. In fact the first option hardly exists, in what would be a Case 
A system if it started with nearly equal masses. 
\par Wilson \etal (1985) and Olson \& Etzel (1994) suggested that the gainer 
is in very rapid rotation, as a result of accretion. If this were {\it uniform} 
rotation, as is assumed in the models computed here, the effect on
the star's radius would be modest, but it is quite likely that there is
substantial {\it differential} rotation, as layers are added that are relatively
rich in angular momentum. Such differential rotation, combined with rapid
(thermal-timescale) accretion can possibly account for the large radius
of the gainer. In fact we may have too much of a good thing, because there
is some likelihood that star 2 may swell enough to fill it own Roche lobe,
fairly soon after the onset of RLOF (Case BR above), even without allowing 
for differential rotation. Fig 3 sketches possible conservative evolution 
of this system starting with parameters ($8.7 + 5.6\Msun; 4.5\thin$d), and 
thus a larger mass ratio (1.6) than either of the two previous
models. It {\it just} avoids contact when the masses are almost equal,
but nevertheless star 2 still has substantially too small a radius at the 
present mass ratio. The evolution is Case BN, but just misses Case BR at equal 
masses, as well as Case BB at the end. We appeal to the accretion
of angular-momentum rich material to account for the observed larger radius. 
It is arguably more important in this case than in the previous two, because
of the assumed higher initial mass ratio. Star 2 has increased its mass by
117\% in our preferred model, whereas the increases were 66\% and 78\% for
V356 Sgr and AQ Cas respectively.
\par It is not clear what will be the long-term outcome of evolution into
contact (Case BR). On the one hand the effect might be minor, with the
system emerging again from contact much as it would have been if contact were
ignored. The enigmatic system $\beta$ Lyr (Wilson 1974) may have done just 
this, if one accepts the evolutionary model of Zi{\'o}{\l}kowski (1976). 
But on the other hand low-mass contact binaries (W UMa systems)
appear to reverse the general direction of mass transfer (supposing they
started with conventional RLOF from star 1 to star 2), and evolve
away from equal masses towards very extreme mass ratios, perhaps until
star 2 is completely merged into star 1. This is discussed at some length
by Eggleton (1996).
\subsection{$\phi$ Per}
\par The next three systems that we discuss are arguably in the detached
post-Algol stage predicted by the previous models. The system $\phi$ Per in Table 1 is 
a binary in which the low-mass component is probably a helium core remnant. The 
emission-line character of the helium spectrum of star 1 suggests that this component 
is losing mass by wind, \ie\ non-conservatively, but apparently not on the scale 
of Wolf-Rayet winds. Since the orbit of $\phi$ Per is nearly circular, it seems a 
plausible candidate for post-RLOF Case B. The present mass of star 1 suggests an 
original mass of $\ts 6 - 7 \Msun$.  This is consistent with the fact that
if the evolution has been conservative, more-or-less, then the original mass 
must also have exceeded $5.2 \Msun$, the average of the two present masses. 
Clearly there is not much scope for systemic mass loss, according to these figures. 
We guess that the original parameters were roughly $(5.9 + 4.6 \Msun; 8\thin$d). 
The predicted mild initial mass ratio is reasonably consistent with 
conservative RLOF.  The resulting evolution is shown in Fig 4.
The above starting model leads by conservative evolution to 
parameters $(1.16+9.3; 115\thin$d) in a late detached stage lasting from $66$ to 
$80\thin$Gyr, with star 1 a small helium-burning remnant. Then star 1 returned 
to the giant branch (Case BB), lost further mass, and the orbit widened further to
parameters ($0.91 + 9.6\Msun; 255\thin$d), before terminating itself after 3000 
timesteps. By that time it was heading rapidly to the CO white dwarf region.
\par In $\phi$ Per, star 2 appears to be in rapid rotation, which is not surprising
since it would have been spun up considerably during RLOF. It will presumably spin 
down in the future, perhaps partly by wind but certainly by evolutionary expansion.
The rotation has a less major effect on the radius of star 2 than in RZ Sct,
perhaps because it is dying away after RLOF ceased. In our model of $\phi$ Per
$M_2$ increased by 102\% during RLOF, not quite as much as star 2 of RZ Sct
in our model above.
\subsection{3 Pup}
\par The system 3 Pup contains a component which is very central in the 
Hertzsprung gap. Although the small mass function might only indicate that 
it is observed at a low inclination, Plets \etal (1995) noted that its 
emission lines are doubled, more suggestive of a ring due to an accretion 
flow seen at high inclination. They tentatively estimated a large mass ratio, 
$\ts 20$, but this estimate depends strongly on where the ring is presumed 
to be. Since this ratio would lead to embarassingly high masses for both 
components, we tentatively assume a more modest ratio of 12, which leads to 
masses of 1 and 12$\Msun$. This further suggests to us that the system is a 
post-RLOF remnant, the unseen component being the remnant of star 1. Such an 
extreme mass ratio ($q\ts 0.083$ in our convention), along with the current period, 
is wholly consistent with conservative Case B RLOF, starting with parameters of
say ($7 + 6\Msun; 4\thin$d). More specifically, with these parameters we find
another Case~BB, with star 1 having a first episode of RLOF which leaves
a helium-buning star, and then a second episode as the helium star expands back 
temporarily to the giant branch. Star~2, the observed supergiant, has itself evolved 
to core helium burning, and might be expected fairly soon to fill its Roche lobe 
and initiate {\it reversed} RLOF. But the existence of emission lines suggests
that some mass transfer is already under way, in the form of accretion from
a wind rather than RLOF. If only a small amount of mass is lost and/or
transferred this way, the RLOF in the future would be expected to be very
rapid (reverse Case BD), because of the extreme mass ratio, and to result in 
a common-envelope episode, spiral-in, and then a short-period binary. We will 
argue, however, that the supergiant, while losing much mass rapidly, will 
{\it not} decrease its period by a large factor, and will instead become a wide 
evolved binary like V379 Cep, which we consider in Section~3.1.
\subsection{HD51956}
\par HD51956 is a system with a well-determined spectroscopic orbit of surprisingly 
short period for an F8 supergiant (Burki \& Mayor 1983). Ake and Parsons (1990)
observed it with IUE and found an early-type (B2 - 3) companion that was 6 - 7 magnitudes
fainter than an MS companion of this type would be expected to be. There is evidence
of accretion activity. This is largely consistent with the interpretation that
this is another post-Algol approaching reverse RLOF. We argue again that the interaction 
of the hot component is with a wind from the slightly-detached F8 supergiant, rather 
than from RLOF itself, since the expected very small mass ratio would make RLOF 
extremely unstable and rapid. The wind may be partly due to the disturbance to 
the supergiant atmosphere caused by the companion in its close orbit. The orbital
period is very reasonable for a Case A system at this late stage in its evolution,
as is the small mass-function. That the mass function is smaller than in 3 Pup may
simply mean that the inclination is somewhat lower, but might also mean that the
masses are somewhat lower too.
\subsection{Provisional Summary}
\par We have suggested that all three systems V356 Sgr, AQ Cas and RZ Sct are
reasonable examples of the semidetached phase of conservative Case A/B evolution, 
although in the third example we have to appeal to a great deal of (non-uniform) 
rotational expansion to account for the large radius of star 2. The further three 
systems $\phi$ Per, 3 Pup and HD51956 are equally reasonable examples of the 
succeeding post-RLOF state in the same kind of evolution. The next four binaries, 
however, are much more problematic.
\section{Some Difficult Cases}
\par It is not easy to prove conclusively that a binary must have started with
more mass than it presently contains, but we believe that the case is very
convincing for the first two systems in this Section, and moderately
convincing in the last two. The strongest case can only be made for a system
which is both double-lined and eclipsing, and only the first example falls in
this category.
\subsection{V379 Cep}
\par V379 Cep (Table 1) is a very extraordinary system. Because it is eclipsing, 
the low masses of the two components, each  $\tls 30\%$ of what we might 
reasonably expect, cannot be explained as a result of a low inclination. It 
apparently contains {\it two} hot under-massive remnants. {\it One} can
be accounted for by legitimate RLOF, as in several systems above, but the other
cannot. We suggest that this is a system that has survived {\it two} episodes 
of RLOF, a fairly conservative one in the forward direction, widening the 
period from a few days to $\ts 100\thin$d, and a second one, highly 
non-conservative of mass but fairly conservative of angular momentum, in the 
reverse direction. 
\par The phrase `fairly conservative of angular momentum', used in the above
paragraph, is deliberately somewhat vague. Obviously a system is unlikely
to lose mass without losing angular momentum (although because angular
momentum is a vector it is not impossible). 
But we do not have a specific model to determine {\it a priori} the amount 
of angular momentum lost. What we mean is that if the system mass decreases
by say $3$, the angular momentum and the orbital period change by comparable 
factors, and not by much larger factors as in the case of common-envelope
evolution. A spherically symmetric wind from either component, if it
somehow fails to interact with the other component on its way out of the
system, will conserve angular momentum per unit {\it reduced} mass (an adiabatic 
invariant), and lead to a period increase: $P\propto (M_1+M_2)^{-2}$. This is 
probably the smallest amount of angular momentum loss that one can plausibly
envisage for a given amount of mass loss, although some contrived situations 
can give even less (well-directed jets, counter-rotating magnetic stars, ...). 
By `fairly conservative' we mean angular momentum loss of this order, and not 
ten or a hundred times more.
\par Our suggestion that the reverse RLOF can be highly non-conservative of 
mass but fairly conservative of angular momentum is rather unorthodox, and 
so we attempt to consider rather carefully the arguments that lead us to this
conclusion for V379 Cep.
In the first place, we conclude that both components must be somewhat
analogous to blue horizontal branch stars, i.e. they must both have 
helium-burning cores surrounded by relatively hydrogen-rich envelopes. We 
know of no other kind of structure which can produce a high luminosity at 
a comparably low mass, high effective temperature and radius, and also be 
reasonably long-lived. Short-lived largely degenerate remnants
of, say, red giants that have recently lost their envelopes might in principle
pass through the region of the HRD where the two components are found, but 
it seems incredible that both should be doing this at the same time, since
they would evolve very rapidly. One might suppose that {\it one} of these
components is doing this, but there is no sign of a recently-ejected planetary 
nebula that would be expected. The horizontal-branch morphology that we suggest 
would be short-lived compared with normal main-sequence stars in the same 
region of the HRD, say $\ts 1\thin$Myr against $\ts 30\thin$Myr, but not 
{\it very} short-lived. 
\par It is a reasonably general
conclusion of stellar evolution that a star spends most of its life either
on the main sequence or else as an inert compact remnant (WD, NS, BH): the
intermediate stages are all fairly short-lived compared with these. 
Furthermore, the lifetime on the main sequence is a strong function of initial mass.
If therefore a binary contains two components which are both in a short-lived
intermediate state, we would normally conclude that the two components were
of rather similar initial mass: probably $1\tle q_0\tls 1.2$. This still applies 
if there is mass transfer in a binary, except that the constraint on $q_0$ is
somewhat weaker. In Paper I we identified a subset
of Case A evolution which we called AL: `late overtaking', analogous to
Case BL defined in the previous Section. In these systems star 1 evolves 
more rapidly than star 2 up to RLOF, but then after substantial
transfer of mass star 2 is able to catch up with and overtake star 1, at any 
rate to the extent that it swells to fill its own Roche lobe before star 1
becomes an inert remnant (WD, NS, BH). This only happens in a specific 
domain of mass ratio and period at given $M_{10}$, as illustrated in pale 
blue in Fig. 8 of Paper 1. The initial mass
ratio must be in the range 1 -- 1.5, and the period in the range 3 -- 5$\PZ$
(straddling the Case A/B boundary), but both ranges themselves depend 
on the mass $M_{10}$: see Fig 9 of Paper I. If the initial mass ratio 
is larger than $\ts 1.5$, or the
period is $\tls 3\PZ$, the system is liable to evolve into contact, and the 
subsequent evolution is unclear. If the period is greater than $\ts 5\PZ$
then generally star 1 is so far ahead in its evolution before RLOF that star
2 has no chance of catching up afterwards. Star 1 will already be a WD or
NS when star 2 reaches its Roche lobe. Thus Case AL/BL seems the most
probable way of achieving the kind of result we want.
\par Consider conservative evolution of a system starting with ($7 + 6.3\Msun,
3\thin$d). This is shown in Fig~5. After RLOF, in Case AL, the parameters were
($1.1 + 12.2\Msun, 100\thin$d). The interval of RLOF was interrupted by
a brief detached phase, at parameters ($2.9 + 10.4\Msun, 9.5\thin$d) when
star 1 was at the end of its core hydrogen-burning phase. The run was terminated
when star 2 as a yellow Hertzsprung-gap supergiant just filled its own Roche lobe,
so that reverse RLOF (not allowed for in the code) was about to begin. Simultaneously 
the core of star 2 had just started helium burning. Also simultaneously, star
2 was in the middle of the Cepheid strip. Star 1 was already a core 
helium-burning star. We believe that both 3 Pup and HD51956 are in rather similar 
states, but slightly earlier in their evolution.
\par Although the usual expectation is that star 2's reverse RLOF will be very
fast, and lead to a common-envelope episode and spiral-in (reverse Case BD), we 
suggest that in order to obtain a system like V379~Cep we need something different. 
If at, or shortly before, star~2's RLOF star~2 ejects almost its entire
envelope to infinity, then (in our specific model) it will leave a core 
helium-burning remnant
of $2.8\Msun$, more or less the mass of the component of V379~Cep which we 
identify as star~2 in Table~1. We suppose in addition that star~1 accretes
a small proportion ($\ts 8.5\%$) of this ejected envelope. This would increase
its mass to the observed value ($1.9\Msun$). 
\par If the envelope of star 2 is almost entirely ejected, in a roughly spherically
symmetric manner, we would expect the orbital period to increase by a
considerable factor. However, we assume that (a) there is some dynamical
friction on star 1 within the expanding envelope, as in classical
common-envelope evolution (Paczy{\'n}ski 1976), (b) the accretion on to
star 1 though much less complete than in conservative evolution acts towards 
{\it decreasing} the period, and (c) therefore the period is more-or-less 
unchanged. This is pure supposition, but it does not seem unreasonable.
\par A reason for supposing that the mass-loss mechanism is very rapid
and erratic (as is seen in P Cygni stars) is that the orbit of V379 Cep 
is eccentric ($e=0.15$). This does not have to mean that all the mass
loss took place in the course of one orbit, but only that the rate, at
least in the final stages, fluctuated on that sort of timescale.
\par An extra observational point that might favour such a scenario of rapid
envelope ejection is that
V379 Cep appears to be in an empty `bubble' near the edge of a star-forming
region (Gordon \etal 1998). Such a bubble might have been created by the rapid 
outflow of much of the envelope, as we hypothesise, perhaps $0.1\thin$Myr ago.
\par It may be a problem that the remnant of $*2$ appears embarassingly normal.
We might expect it to be a Wolf-Rayet-like object, since Wolf-Rayet stars
are often identified with the helium cores left after a supergiant has lost
its envelope in a P Cygni wind. However we are dealing here with substantially
less massive remnants, from somewhat less luminous supergiants. We would
however expect that star 2 might show the nitrogen enhancement of an OBN
star (like V2174 Cyg below), as should star 1.
\par What might be the mechanism of such envelope ejection? We leave that to
Section 3.5, where we suggest three rather different mechanisms. One is related to the 
Humphreys-Davidson limit as potentially modified by a close companion, one to 
Cepheid pulsations also modified by a close companion, and one to the influence of 
eccentricity in the orbit prior to RLOF. The last is not very likely in the
present system, since we would expect the orbit to have been very highly circular
after the first episode of (forward) RLOF, but it might be important in the three
other systems we consider below.
\subsection{$\upsilon$ Sgr}
\par The system $\upsilon$ Sgr, containing a hydrogen-deficient A supergiant of
low mass, is a surprisingly bright member of the rather rare class of 
hydrogen-deficient star. It has long been known as a single-lined spectroscopic 
binary, but Dudley \& Jeffery (1990) were able to detect a weak secondary 
spectrum in the UV. Although the system does not eclipse, there are faint indications 
of variable H$\alpha$ absorption round the orbit (Nariai 1967), suggestive of an
accretion disc and therefore of a fairly high inclination. Thus we may accept that
the masses are not very different from the values of $M\sin^3i$ listed in Table 1.
\par At first one might suppose that this is a Case A or more probably
B system, with star 1 stripped down to near its core by one or two episodes 
of RLOF in the forward direction. But this is quite a difficult model to support 
because the mass ratio should be {\it much} more extreme: $q\ts 0.08 - 0.16$,
rather than 0.63. Note that this conclusion is quite independent of the uncertain
inclination. This therefore appears to be another case where much mass has been 
lost from the system as a whole, but not much angular momentum. We propose initial 
parameters of ($\ts 10 + 3\Msun; \ts 150\thin$d). Star 1, on or shortly before reaching 
its Roche lobe, dropped rapidly from about $10\Msun$ to its terminal main-sequence 
core mass of $\ts 2.5\Msun$, while star 2 accreted little: about 13\% of the 
ejected envelope. The period would have increased by quite a large factor if all 
of the envelope of star 1 were ejected isotropically to infinity, but as with 
V379 Cep we suppose that this was mitigated by both hydrodynamic drag of star 2 
in star 1's expanding envelope, and by partial accretion, leaving the period much 
the same as originally. 
\par Our evolution code does not yet include a model of this rapid envelope ejection
process, but we attempted to approximate at least the evolution of star 1 by putting
it in an initial binary with parameters ($10 + 8\Msun; 23\thin$d). Star 1 would have 
rapid (thermal-timescale) but conservative RLOF, and end up in a binary with something 
like the same period as observed in $\upsilon$ Sgr. We expect that this transition 
(for star 1 only) might not be very different from the transition by way of  
rapid envelope ejection in a binary with initial period something like the current 
period. The remnant of star 1 was $2.52\Msun$. It shrank during core helium-burning 
to $\ts 0.6\Rsun$, but then expanded at a late stage back to giant radius. It was 
roughly an AI supergiant when it almost simultaneously reached carbon ignition and 
its Roche lobe (like V356 Sgr). If we accept this at face value, then $\upsilon$~Sgr 
should be increasing its radius on a timescale of $\ts 1300\thin$yr, and may be 
within a few centuries of a supernova explosion.
\subsection{SMC 0045-7319}
\par The SMC system 0045-7319 (Kaspi \etal 1994) is a pulsar binary which is unusual 
because it contains both a B star and a {\it radio} pulsar, with no sign of the 
accretion (from stellar wind) that such massive binaries with neutron stars commonly 
exhibit in the form of variable X-ray flux. The companion seems to be an unusually 
{\it in}active B star, without any Be characteristics. This has the beneficial effect 
that the pulse period is hardly erratic at all, and so the parameters of the orbit 
derived from the pulses are extremely sharply defined: a speeding up of the 
{\it orbit} on a timescale of $0.5\thin$Myr, presumably due to tidal friction, is 
measurable. The masses given in Table 1, from Bell \etal (1995) and Kaspi \etal (1996),
combine (a) the very accurate pulsar mass function, (b) the much less accurate
mass function of the B star, and (c) the hypothesis that the neutron star has
mass $1.4\Msun$. These also lead to an inclination of $\ts 45\deg$. The radius
of the B star comes from its luminosity, a temperature appropriate to its spectral
type, and the known distance to the SMC.
\par The system presents some substantial problems regarding its evolutionary status
(van den Heuvel \& van Paradijs 1997, Iben \& Tutukov 1998). It cannot be the result
of reasonably conservative RLOF, followed by a helium-star phase and then a supernova, 
because in such evolution star 2 ends up substantially more massive than the {\it original} 
star 1, as illustrated by V356 Sgr and AQ~Cas in Section 2. The present mass
($8.8\Msun$) of star 2, though uncertain, is too low: star~1 in this scenario 
would not have been massive enough to have a supernova explosion.
On the other hand we cannot have had a route which involves a normal P Cygni phase
of mass loss and no RLOF, because this would require a quite massive star 1, say 
$\tgs 30\Msun$ originally, and in that case the much lower-mass star 2 should not
have reached its surprisingly large (though also uncertain) radius, as illustrated
by the value of $Y$ well in excess of unity. We feel that in
order to model this system we have to invoke the same mode of rapid ejection referred
to above: a drastic loss of mass from star 1 when star 1 was close to, but perhaps 
not quite at, RLOF, in a fairly wide orbit, {\it without} a drastic shrinkage of 
the orbit. We require that star 1 was only about $10 - 11\Msun$ originally, in order 
for star 2 to show as much evolution as it does. With initial parameters ($10.5+8.8\Msun; 
\ts 30 - 150\thin$d), and assuming no significant mass {\it transfer} (as we are 
driven to), star 2 would have radius $6.5\Rsun$ when star 1 reaches its supernova 
explosion at age $24.0\thin$Myr.
\par We cannot be very confident about the original period, because the supernova
explosion would have changed it in a way that is hard to predict, at least if the
explosion were asymmetric as is generally thought to be the case. However, even
with an asymmetric kick we can say that the original orbit must have had a semimajor
axis somewhere between the periastron and apastron separation of the current
orbit, corresponding to a period between about $5\thin$d and $150\thin$d. At the
shorter end of this range the system would have been not unlike V356 Sgr, AQ cas
and RZ Sct above, which appear to have been subject to fairly normal Case A/B RLOF.
Consequently we opt for the upper end of this period range, making the system
more similar to V379 Cep before its reverse RLOF, which we have argued was
very non-conservative of mass.
\par Although for such stars in an orbit of say $ 30 - 100\thin$d the classical 
expectation would still be of fairly well-behaved Case B RLOF, and the Humphreys-Davidson 
limit would be an irrelevance, we argue here, as with $\upsilon$ Sgr and V379 Cep
(above) and V2174 Cyg (below), for an alternative with rapid substantial mass loss 
and little shrinkage at a modified Humphreys-Davidson limit where the presence of
a Roche lobe brings down the luminosity criterion for instablity. 
\par It might be suggested that the evolutionary growth in radius of star 2 has 
taken place {\it after} the supernova explosion. However, several Myr would be 
required to make a
significant difference. The model of tidal friction by Eggleton \& Kiseleva-Eggleton (2001), 
when integrated backwards in time, suggested that $\ts 1\thin$Myr ago the orbit 
had $P\ts 200\thin$d, $e\ts 0.92$. This is not out of the question given that the 
second and only other known {\it radio}-pulsar high-mass binary (1259-63, SS2883; 
Johnson \etal 1994) has a period of $1237\thin$d and $e=0.87$, but it is difficult 
to imagine the age being extended to a significant fraction of $24\thin$Myr.
\par We may be able to avoid the conclusion that this is a difficult system,
analogous to the other three in this Section, if we (a) push up the mass of
star 2 to its highest value consistent with observational uncertainty, perhaps
$12\Msun$, and (b) reduce its radius to the minimum allowable, perhaps $5\Rsun$.
Then star 2 would be substantially less evolved, and therefore star 1 could have been
{\it considerably} more massive originally, say $30 - 50\Msun$. In that case star 1
might have reached the Humphreys-Davidson limit without any help from star 2.
We are pushing the error bars to their limits, however.
\par In estimating radii and luminosities at a given mass we have not made
allowance for the fact that the SMC is somewhat metal-poor compared with
young stars in the solar neighbourhood. We believe this will not be a major factor,
and is liable to point in the direction that star 2 must be somewhat more
evolved still, in order to have reached its estimated radius.
\subsection{V2174 Cyg}
\par V2174 Cyg (Table 1) is an OBN star (Walborn 1976). Bolton \& Rogers (1978) 
showed that several OBN stars are in binaries, and argued that the 
nitrogen-richness of this class is a consequence of RLOF: the loser's core has 
been exposed to the point where CNO processing has enhanced the nitrogen abundance. 
Most OBN binaries listed by Bolton \& Rogers (1978) have periods of a few days, 
but V2174 Cyg has a long period, so that if binarity was important star 1 must 
have evolved far into the Hertzsprung gap, and then retreated. 
\par There appear to 
be at least four main possibilities, in principle, for the visible component: 
\sep (a) it is star 1, but has lost mass {\it via} conservative RLOF to a 
companion, exposing an N-rich core; 
\sep (b) it is star 2, which has gained N-rich material from the companion, now 
invisible, which earlier filled its Roche lobe; 
\sep (c) it is star 1, and has suffered some kind of internal mixing unrelated 
to star 2, or
\sep (d) it is star 1, but has lost mass by single-star, or 
binary-enhanced, wind, leaving the companion fairly unaffected. 
\pn All of these 
alternatives have problems. In (a), star 2 would now presumably be the more 
massive component, perhaps by a factor of 4 or 5, so it is surprising that it 
is not seen: most Algols or post-Algols are dominated by the light of star 2. In 
(b), star 2 would now be the less massive star by the same sort of factor, in 
which case the large mass-function implies colossal masses for both 
components. In (c), the lack of connection with star 2 seems contrary to
Bolton \& Rogers's (1978) observation that most or all OBN stars are in binaries;
and it is also quite difficult to understand the large mass function if the
BNI star, being more luminous, is the more massive. In (d), star 1 can be less 
massive, but even if it has half the mass of star 2, star 2 would still have to 
be $\tgs 13 \Msun$, and so might be expected to be visible. 
\par We suggest that the least unsatisfactory model is (d), for which we 
invoke the same process of `envelope ejection'.  The BNI precursor, star 1, was 
originally quite massive, say $\ts 20\Msun$, with a companion of $\ts 12\Msun$. 
It evolved into the Hertzsprung gap, at luminosity $\ts 10^5\Lsun$, and would
have passed some considerable way across it. But triggered by its approach to 
its Roche lobe, or by the start of RLOF, in a roughly $100-500\thin$d orbit, star 
1 became unstable like a P Cygni star and ejected its envelope, leaving a remnant 
of $\ts 6.5\Msun$. We imagine that, in analogy with V379 Cep, there might have 
been partial accretion by star 2 of some fraction of the mass lost, so that star 
2 is now $13\Msun$, with $L_2\ts 10^{4.1}\Lsun$). The period, also by analogy, 
was not much altered -- perhaps only by a factor of two. The core has just enough 
of an H-rich envelope to be in the middle of the Hertzsprung gap, perhaps like 
the precursor to SN 1987A (McCray 1993), rather than to be a hot WR-like object. 
Star 1 approached near enough to its Roche lobe to lower the eccentricity to a 
modest value, but not so much as to lower it to zero; or alternatively the mass 
loss was sufficiently erratic that it created the present eccentricity. The 
remaining masses would be consistent with the mass function, while the 
luminosities, particularly in the visible region of the spectrum, could still 
differ by a factor of $\tgs 8$ and so explain the invisibility of star 2.
\subsection{Provisional Summary}
\par The four systems above apparently have very little in common. However,
they all are difficult to fit within the context of {\it conservative}
evolution of Case B systems, unlike the six systems dicussed in the previous
Section. At the same time, they all have longish, but no very long, periods, 
and appear to have come from a mass range of $\ts 10 - 20\Msun$. We believe 
that, taken together, they suggest a mode of binary evolution that is highly
non-conservative of mass, but which does not involve drastic shrinkage of
the orbit.
\section{Mechanisms for Rapid Envelope Ejection}
\par We mentioned in Section 1 that a star of
$\ts 30 - 40\Msun$ apparently loses all its envelope spontaneously
as a yellow supergiant. A {\it single} star of less mass does not,
but it may well lose at least a part of its envelope spontaneously.
We hypothesise that the presence of a binary companion helps to
make what would be a reasonably stable distended atmosphere substantially
less stable. This could be through tidal friction, which would be
attempting to spin up the envelope substantially as it expands towards
its Roche lobe; left to itself, the envelope would be spinning {\it down}
substantially.
\par The essential mechanism of mass loss at the Humphreys-Davidson limit is 
itself unclear.
It is usually conjectured to be a result of intrinsic envelope instability,
perhaps due to the fact that the luminosity of such stars is close to
the Eddington limit. Radiation pressure is generally thought to accelerate
the wind to the rather high terminal velocities seen in P Cygni stars, which
are mass-lossy supergiants at or near the Humphreys-Davidson limit; radiation 
pressure is of course the major component of pressure at the Eddington limit. 
We conjecture that the AIab and FIb supergiants of 3 Pup and HD51956 may be 
sufficiently close to the AIa and FIa supergiants near the Humphreys-Davidson 
limit that the effect of also being near to their Roche lobes
may push them closer to the instability of
P Cygni stars. It is not necessary that the mass loss be instantaneous,
only that it be rapid enough to cause total envelope loss at or shortly
before RLOF. Indeed, it might be that RLOF by itself can stimulate these
supergiants to behave like somewhat more luminous supergiants, and get
rid of their entire envelopes at sufficiently high velocities that little
is accreted by star 1.
\par Even if the above mechanism is physically reasonable, it is still
somewhat uncertain that it will allow the kind of precursor that we
suggest to evolve into a system just like V379 Cep. We might expect
that at least one of the two components would be a rather Wolf-Rayet-like
remnant, since Wolf-Rayet stars are thought to be the remnants of mass
loss at the Humphreys-Davidson limit. We would argue that the addition 
of a relatively slight hydrogen-rich envelope to the remains of star 1, 
as we hypothesised above to explain the fact that the orbital period is 
not much greater still, might make it more like
a horizontal-branch star than an almost pure helium-main-sequence star,
and that the exposed core of a $12\Msun$ star (star 2) might retain
more hydrogen-rich envelope than the exposed core of a $40\Msun$ star,
and thus also be less Wolf-Rayet-like. It is amazing that two stars
that have had the kind of adventures we hypothesise should end
up looking exactly like ordinary main sequence stars, but this must
be the case whatever the mechanism. However we expect that the
remnant would show an enhancement of nitrogen, as in V2174 Cyg above,
because of the exposure of CNO-processed material.
\par A second mechanism that we consider is based on the fact that
at least some of the systems we are considering here have stars that should
be reaching their Roche lobes at about the same time that they should
be reaching the Cepheid strip. It may be coincidence that in our 
model above for V379 Cep star 2 is just in the instability strip as it 
reaches its Roche lobe (Fig 5a). But it does not seem improbable that the
inherent instability of stellar envelopes in the Cepheid strip might be
considerably amplified by the presence of a close companion. 
\par We feel that we can be relatively confident about the period of
V379 Cep immediately before its episode of reversed RLOF, because the
evolutionary state of both components seems to demand that it started
its RLOF near the Case A/B borderline. We are less clear about the
initial periods in the other three difficult systems, because there
is no such guidance, and we simply assume, by analogy with V379 Cep,
that the period was not much changed by the highly non-conservative
episode. But the periods we assume do in practice put each loser
close to the Cepheid strip when RLOF would have been reached: a
somewhat larger period in the case of the most massive system,
V2174 Cyg.
\par Several Cepheids are known to be in binaries, including some where
the orbital period has been determined spectroscopically. The shortest
orbital period that we know of is S Mus (Table 1). The radius
of the Cepheid in this system is probably about a quarter of the Roche-lobe
radius, and must have been larger still, by about a factor of three, at 
the earlier point in its evolution when core helium ignited. There is
no evidence here that the pulsations are converted into dramatic
outflow by the presence of the Roche lobe, but then there is presumably
a major difference between filling 25\% of the lobe (by radius) and
100\%. The eccentricity of S Mus is quite modest ($e=0.08$), which would
be consistent with the supposition that it was briefly  near but not
extremely near to its Roche lobe at helium ignition.
\par A third factor that might influence the nature of RLOF in three of 
our four difficult binaries is the eccentricity of the orbit before RLOF 
begins. Our model of tidal friction (Paper II) allows us to make an estimate 
of the time it takes to circularise an orbit. For Case A and very early
Case B systems this is less than the time to evolve to RLOF, because
the star spends a large part of its nuclear lifetime at a radius of
$\tgs 20\%$ of its Roche-lobe radius, when tidal friction is
important. But if the period is sufficiently long that the star
expands by a factor of several on its rapid (thermal timescale) evolution 
across the Hertzsprung gap, then we find that most of the circularisation
takes place only when the the star is $\tgs 80\%$ of its Roche radius.
This may not be sufficient to prevent some kind of pulsed mass transfer
which we hope might help to promote mass loss to infinity.
\par In addition to any of the above three factors is the fact that the
star will be rotating rapidly, compared to a single star of the same
size. The timescale
for synchronisation is a good deal shorter than the timescale for
circularisation, and we usually find that the spin rate is within a
factor of two of the orbital rate well before RLOF, even when the star
is expanding on a thermal timescale. We feel that the most likely agent
of rapid envelope ejection is the combination of rapid rotation (relative
to single stars of the same size) and proximity to the Humphreys-Davidson
limit, but it would not be surprising if both Cepheid instability and
eccentricity tended to enhance the efficiency with which the star's
luminous energy flux can be channeled into kinetic energy flux.
\par The thermal, or equivalently Kelvin-Helmhotz, timescale of a star
is just the timescale of mass loss {\it if} luminous energy is converted
into outflowing kinetic energy with $\ts 100\%$ efficiency. Something like
this is presumably achieved at the Humphreys-Davidson limit, where
$L/M$ in solar units is $\ts 10^4$. The efficiency evidently drops
off rapidly as $L/M$ decreases. In our moderately wide binaries 
($P\ts 50 - 250\thin$d) we require rapid ejection even at 
masses of about $10\Msun$, where $L/M$ has dropped by about a factor
of 10. We might expect the efficiency for single (slowly rotating)
stars of the same size to have dropped by say $\ts 10^3$, but if
rapid rotation, perhaps combined with Cepheid instability and
eccentricity, mitigates this factor to say $30$ we would still 
have mass loss which is virtually instantaneous compared with the 
nuclear evolutionary timescale.
\par Common-envelope evolution is thought to be an efficient means
of losing mass, but relies on {\it orbital} energy to drive the mass
loss. However one cannot extract orbital energy without extracting
orbital angular momentum as well: indeed one wants angular momentum to
be lost in abundance, in order to explain various types of very close
binary (e.g. low-mass X-ray binaries and cataclysmic binaries) that
were almost certainly much wider binaries at an early stage of their 
evolution. However the problem systems that we identify in Section 3,
though they will have lost angular momentum with mass, do not appear
to have shrunk their orbits very significantly. This is particularly
true of V379 Cep, where we feel driven to a fairly specific zero-age
model, and from there to a period just before reversed RLOF that is
(no doubt somewhat coincidentally) almost exactly the same as is seen 
after the event.
\par It would not be fair to leave this discussion without noting that
the binary AX Mon (Table 1) may be telling a different story.
If interpreted as a conservative Case B product, our `canonical' starting
parameters would have been ($10.4+7.7\Msun; 95\thin$d), not very different
from our supposition regarding 0045-7319. However one can also appeal to
an almost entirely non-conservative model in which the starting parameters
might have been ($14+13\Msun; \ts 200\thin$d), and presumably to a range
of possibilities in between. 
\par Elias \etal (1997) point
to a rate of period {\it decrease}, significant at the 3$\sigma$ level,
on a timescale of $1.7\times 10^4\thin$yr. The sign is unexpected
whichever of the above two models one attempts to support. It presumably 
means that there is disproportionate angular momentum loss, which may 
indicate stellar wind (to infinity) along with magnetic braking. Both the 
wind and the magnetic field might be attributed to dynamo activity in the 
K giant's convective atmosphere, much as we have attempted to model 
elsewhere (Paper II) for more normal Algols of shorter period.
\par AX Mon is a single-lined, non-eclipsing system, and so the
parameters quoted in Table 1 are based on a substantial amount of
inference. One of these appears to be the {\it assumption} that the
K star fills its Roche lobe. This is perhaps a little surprising for
the luminosity class III ascribed to the K star. Early K giants are
often found in binaries with periods as low as $10 - 60\thin$d. There
are clear signs of gas streams in the system, but it is not impossible
that wind from the K star might interact with the companion to produce
effects not unlike those associated with RLOF.
\par For the present, we do not feel we can draw any firm conclusion
from AX Mon. But we would emphasise the general principle that it is not
good enough to invent a process that accounts for some systems, while
ignoring the process to account for some others that should otherwise
be similar.
\par There are several more apparently interactive binaries with periods
comparable to AX Mon's, and to our `difficult' systems, but most or all
of them can be interpreted as systems of substantially lower total mass.
Thus they do not illuminate the question of whether stars of $\tgs 10\Msun$
may eject much mass to infinity if they are late-B/A/F supergiants with a 
close companion.
\par The orbital period necessary for our ejection mechanism can probably not be 
much {\it greater} than the values we have inferred, because there exist a handful 
of K/M supergiant binaries in or near to the mass range ($\tgs 10 - 30\Msun$) we 
are interested in which have longer periods. We mention V1488 Cyg, V695 Cyg, VV Cep 
and AZ Cas (Table 1). The first two, which are eclipsing double-lined systems,
are reasonably well modeled by normal evolution 
without any mass loss (Schr{\"o}der \etal 1997). They are slightly less massive 
than the range in which we are primarily interested, as well as of substantially
longer period. One or both of these factors appears to mean that they have
escaped envelope ejection. The last two are wider still. They are only single-lined, 
but with large mass-functions that would indicate masses of at least 20 and 14$\Msun$ 
for the two M supergiants {\it if} they are more massive than their early-type 
companions. As far as we can tell the masses quoted are based on the {\it conjecture} 
that the mass ratio is $\tgs 1$. But if on the other hand these
M supergiants have lost significant mass, so that they are now the less massive
component, then we cannot say very much about either their current masses or their
original masses. But at least we can say that all four primaries have avoided
going directly from mid-Hertzsprung-gap supergiants to hot remnants, as we
require for our ejection process, and instead have evolved to much larger radii.
This seems to confirm that we need a close enough companion to an A/F supergiant
to effect the ejection.
\section{Discussion}
\par We have seen that some massive and fairly close binaries, the first 
group of Table 1, can be accounted for reasonably well by conservative RLOF 
in Case A or very early Case B, and that some other, somewhat wider, binaries 
-- the second group -- can be seen as later stages of similar systems. But 
four systems at least (the third group in Table 1), which were probably 
substantially wider than the first group but comparable to the second group 
when RLOF began, are hard to account for with the same model. Note that we 
are not distinguishing here between the first episode of RLOF, in the forward 
direction, and the second, in the reverse direction: in our preferred model 
of V379 Cep the first occurred at a short period, and was conservative, but 
the second and non-conservative one was at a much longer period. In our 
preferred models of the other three difficult systems the {\it first} RLOF 
occurred at a fairly long period. A highly non-conservative model is required 
for these, but one which, unlike common-envelope evolution, does {\it not} 
shrink the orbital period down to a day or less. Although our difficult group 
only come from the mass range $\ts 10 - 20\Msun$, we expect that it will 
extend to $\ts 30\Msun$.
\par What we believe distinguishes the difficult group from the first group
and also the last group in Table 1 is their periods, with the implication 
that if a massive star which is about to fill its Roche lobe has evolved far 
into the Hertzsprung gap, but not so far as to reach the red (super)giant 
branch, then it is prone to lose almost its whole envelope
to infinity rather than to its companion. This might be because the 
Humphreys-Davidson limit for single stars is in effect lower for binary stars
with close companions -- close, that is, once the star has expanded to
supergiant radius -- or it might be that pulsational instability of the Cepheid
variety is much more vigorous in a star with a close companion (in the same
sense). It might even be a combination of the two.
\par Possibly the mass ratio immediately before the ejection episode is
also important. However, we seem to have a substantial spread. For 0045-7319
the mass ratio of 1.2 that we hypothesise is rather mild, while for V2174 Cyg,
$\upsilon$ Sgr and V379 Cep the values are 2, 3.3 and 11. Four cases hardly
make a good statistical base, but there is no evidence here that mass ratio
is an important discriminant between our envelope ejection process on the one
hand and conservative RLOF or common-envelope evolution on the other. 
\par Case B is traditionally divided into two major subtypes, early or
late, where star 1 has a radiative or a convective envelope respectively at 
the onset of RLOF. We conclude that several more subtypes are necessary.
There appears to be a significant difference between `very early Case~B'
and `moderately early Case~B', with a division that may correspond to
whether a star is on the left and below, or on the right and above, some
boundary that goes through the middle of the Hertzsprung gap. The first
group may have fairly conservative early Case B, and the second a version
that is highly non-conservative of mass but relatively conservative of
angular momentum. We cannot be very precise about this boundary, but stars 
which have spectral types $\ts$AI/FI/GI, and have masses of $\ts 10 - 30\Msun$, 
as they approach their Roche lobes appear to be in the second category. 
Within the first category, we should remember that even conservative RLOF 
has several subtypes, which we have identified as Cases BD, BR, BL, BB and 
BN; the second category we call BU (`B -- unusual'), for the present. Case 
BD, which has a large mass ratio, probably leads to common-envelope evolution, 
and quite possibly a complete merger rather than a short-period remnant. The 
fate of BR systems is very unclear, but we suspect they also end up as merged 
single stars, by a slower and somewhat more conservative process. Cases 
BL -- BN lead to wider detached binaries, and then to {\it reversed} RLOF, 
which we conclude will usually be Case BU (rev).
\par Case C is presumably similar. In the upper part of our range of masses
stars may ignite helium well before they have crossed the Hertzsprung gap,
so that there will be early Case C as well as late Case C systems. There
will however probably not be `very early Case C' systems, since helium does not
usually ignite before the star is about half-way across the Hertzsprung gap.
We believe that what we have concluded about `moderately early Case B'
will apply to `moderately early Case C', and in effect to most early
Case C systems.
\par We may have to go to {\it late} Case B (or late Case C) for the
kind of systems which undergo common-envelope evolution with spiral-in
to short but non-zero periods, producing low-mass X-ray binaries at high
(star 1) mass and cataclysmic variables at low mass. But such evolution
is not guaranteed; we suspect that it depends strongly on mass ratio,
unlike case BU. We hope to show in a future paper that short-period
highly-evolved systems require not just a convective envelope in the
loser, but a mass ratio $q\tgs 4$ as well. Systems with lower mass
ratios seem, like Case BU, to suffer much mass loss with little orbital
shrinkage.
\par If we are right about the existence of Case BU, then there appear to 
be implications for various
evolutionary scenarios. For example low-mass X-ray binaries are often
conjectured to arise from common-envelope evolution with spiral-in. We start
with a binary containing an OB star and a GK dwarf, and require that when
the OB star has developed  a helium-burning core and a large envelope the
GK dwarf spirals into the envelope and ejects it, reducing the period from
hundreds of days to less than a day. The helium-burning core in this close
binary then evolves to a supernova explosion. But if the initial period is
such that star 1 reaches its Roche lobe somewhat before it reaches
the red (super)giant branch, we would claim on the evidence presented above
that the period does {\it not} shrink drastically. We may need substantially
wider initial binaries (late Case B/C), where star 1 has room to develop a 
deep convective envelope, to do this.
\par The DJEHUTY project at Lawrence Livermore National Laboratory is an
attempt to model stars, including both hydrodynamic and radiation-transport
processes, with a fully 3-D grid. We believe this will be the best way to
investigate the interaction of a lobe-filling A/F/G supergiant with its
companion. Such modelling would allow the possible effect of Cepheid
instability to be included.
\par This work was undertaken as part of the DJEHUTY project at LLNL.  Work 
performed at LLNL is supported by the DOE under contract W7405-ENG-48.

\begin{figure}
\centerline{\psfig{figure=f1.epsi,height=3.0in,bbllx=30pt,bblly=520pt,bburx=570pt,bbury=780pt,clip=}}
{ Fig 1 -- Possible evolution of V356 Sgr (Case BN). Star 1 is the darker line, star 2 the lighter. 
(a) The theoretical Hertzsprung-Russell diagram. (b) Log radii as functions of mass; 
the Roche-lobe radius of either star is shown as the near-parabolic curve. (c) Log radii as functions of 
time. Both stars start near the centres of (a) and (b), on the ZAMS. The run terminates
when star 1 is in the process of igniting carbon, and is therefore on the 
verge of a supernova explosion. It is also very close to filling its Roche lobe again.
Star 2 is still in the main-sequence band at this point.}
\end{figure}
\begin{figure}
\centerline{\psfig{figure=f2.epsi,height=3.0in,bbllx=30pt,bblly=520pt,bburx=570pt,bbury=780pt,clip=}}
{ Fig 2 -- Possible evolution of AQ Cas (Case AN). The panels are the same as in Fig 1.
Because RLOF began shortly before the end of the main sequence, the `hook' due to central
hydrogen exhaustion is just below the middle right. At the end of its evolution star 1 is on the
verge of a supernova explosion. Star 2 is still in the main-sequence band.}
\end{figure}
\begin{figure}
\centerline{\psfig{figure=f3.epsi,height=3.0in,bbllx=30pt,bblly=520pt,bburx=570pt,bbury=780pt,clip=}}
{ Fig 3 -- Possible evolution of RZ Sct (Case BN). The panels are the same as in Fig 1.
Star 2 just missed filling its own Roche lobe (by 3\%) when the masses were nearly equal;
a larger initial mass ratio or shorter initial period would have led to contact at this
point. The observed star 2 (at $M_2\ts 11.7\Msun$) is three times larger than our
theoretical model, perhaps because of recent accretion of rapidly rotating material. }
\end{figure}
\begin{figure}
\centerline{\psfig{figure=f4.epsi,height=3.0in,bbllx=30pt,bblly=520pt,bburx=570pt,bbury=780pt,clip=}}
{ Fig 4 -- Possible evolution of $\phi$ Per (Case BB). The panels are the same as in Fig 1.
The present star 1 is presumably near the helium-burning main sequence, at the LH side
of the HRD. Future evolution of star 1 involves a second episode of RLOF. The run terminates
at 3000 timesteps with star 1 about to settle down as a white dwarf of $0.91\Msun$, and
star 2 still in the main-sequence band. }
\end{figure}
\begin{figure}
\centerline{\psfig{figure=f5.epsi,height=3.0in,bbllx=30pt,bblly=520pt,bburx=570pt,bbury=780pt,clip=}}
{ Fig 5 -- Possible evolution of V379 Cep (Case AL). The panels are the same as in Fig 1.
While star 1 is on or near the helium main sequence, star 2 evolves beyond the hydrogen
main-sequence band, igniting helium in its core very shortly before filling its own Roche
lobe, where the run terminates. Simultaneously it reaches the Cepheid strip. If at this 
point star 2 blows off most of its hydrogen-rich envelope, while star 1 accretes a small 
portion of it, we might expect two horizontal-branch-like stars somewhere near the 
main-sequence band.}
\end{figure}
\end{document}